\documentclass[conference]{IEEEtran}
\usepackage[utf8x]{inputenc}

\usepackage{cite}

\ifCLASSINFOpdf
  \usepackage[pdftex]{graphicx}
  \graphicspath{{./figures/}}
\else
  \usepackage[dvips]{graphicx}
  \graphicspath{{./eps/}}
\fi
\usepackage{enumerate}

\usepackage{array}

\usepackage{tabularx}

\ifCLASSOPTIONcompsoc
 \usepackage[caption=false,font=normalsize,labelfont=sf,textfont=sf]{subfig}
\else
  \usepackage[caption=false,font=footnotesize]{subfig}
\fi

\usepackage[]{footmisc}
\usepackage{url}

\begin{document}
%
\title{Virtual Traffic Lights: System Design and Implementation  }

\author{\IEEEauthorblockN{Rusheng Zhang \IEEEauthorrefmark{1}\IEEEauthorrefmark{3}, Frank Schmutz \IEEEauthorrefmark{2}\IEEEauthorrefmark{3}, Kyle Gerard \IEEEauthorrefmark{2}\IEEEauthorrefmark{3}, Aurélien Pomini \IEEEauthorrefmark{2}, Louis Basseto \IEEEauthorrefmark{2}, Sami Ben Hassen \IEEEauthorrefmark{2}, \\ Akihiro Ishikawa \IEEEauthorrefmark{1}, Inci Ozgunes \IEEEauthorrefmark{1}\IEEEauthorrefmark{3}, and Ozan Tonguz \IEEEauthorrefmark{1}\IEEEauthorrefmark{3}}\\
\IEEEauthorblockA{\IEEEauthorrefmark{1} Department of Electrical and Computer Engineering,
Carnegie Mellon University,
 Pittsburgh, PA 15213-3890, USA}

\IEEEauthorblockA{\IEEEauthorrefmark{2}School of Computer and Communication Sciences, Ecole Polytechnique Federale de Lausanne (EPFL), Lausanne, Switzerland}

\IEEEauthorblockA{\IEEEauthorrefmark{3}Virtual Traffic Lights, LLC,
5503 Covode St.,
Pittsburgh, PA 15217,
USA}
}


%


\maketitle

\vspace{-0.5in}

\begin{abstract}
Traffic congestion is a daunting problem that is affecting the daily lives of billions of people across the world. Recently, a promising new traffic control scheme known as Virtual Traffic Lights (VTL) has been proposed for mitigating traffic congestion. VTL is an infrastructure free traffic control scheme that leverages the presence of Vehicle-to-Vehicle (V2V) communications. Such infrastructure free scheme has several benefits, such as alleviating traffic congestion; reducing the large cost of traffic lights and traffic control systems; reducing carbon emission, etc. This paper reports a DSRC-based prototype design effort on VTL using Dedicated Short Range Communications (DSRC) technology. The experiments performed show the feasibility of implementing VTL with DSRC technology. Preliminary results of the field tests conducted in Pittsburgh with vehicles using VTL equipment indicate that VTL is capable of coordinating traffic at intersections and reducing the commute time of people.
\end{abstract}
\vspace{0.1in}
\emph{keywords: V2V communications, vehicular ad-hoc networks, intelligent traffic lights, intelligent transportation systems}

%
\IEEEpeerreviewmaketitle

\section{Introduction}
\footnote[1]{The research reported in this paper was funded by King Abdulaziz City of Science and Technology (KACST), Riyadh, Kingdom of Saudi Arabia}Traffic congestion is a daunting problem that is affecting the daily lives of billions of people around the world. 

With the advent of Dedicated Short-Range Communication (DSRC) radios being installed in modern vehicles, an infrastructure free intersection coordination scheme, known as Virtual Traffic Lights (VTL) has been introduced as a viable alternative solution to traffic management at intersections \cite{ferreira2010self}. By deciding the right-of-way in a distributed fashion at an intersection through Vehicle-to-Vehicle (V2V) Communications, VTL technology can manage the traffic at an intersection in a self-organized manner.

As a viable self-organizing traffic control scheme, VTL can eliminate the need for infrastructure-based traffic lights which are expensive to install and maintain and are susceptible to failure during natural disasters. Extensive simulations have shown that VTL technology can reduce daily commute time of urban workers by more than 30\%. Different aspects of VTL technology, including algorithm design, system simulation, deployment policy, and carbon emission have been studied by different research groups in the last few years.\cite{ferreira2010self,neudecker2012feasibility,ferreira2012impact,nakamurakare2013prototype,viriyasitavat2013accelerating,bazzi2014distributed,hagenauer2014advanced,tonguz2014implementing,yapp2015safety,bazzi2016distributed,tonguz2016self}. 
	
In this paper, a prototype system design and the performance of VTL technology are reported. Based on experimental results and observations reported in this paper, we have verified the viability of VTL technology as a potential solution for mitigating traffic congestion and reducing the commute time of people during rush hours and non-rush hours.

\vspace{-0.1in}
\section{Related Work} VTL was first proposed by Ferreira et al \cite{ferreira2010self}. In this work, a detailed system design is proposed, together with city-level traffic simulations. The authors point out that currently only around 20\% of all the  intersections in most cities are signalized due to the large cost of traffic signals. The deployment of VTL, which, in principle, can turn all intersections into signalized intersections, will reduce the average commute time by more than 30\%.

The scheme was further improved and designed in more detail by several researchers. Neudecker et al. studied a VTL solution that can handle non-line-of-sight (NLOS) situations \cite{neudecker2012feasibility}. Bazzi et al.  designed a communication algorithm for VTL based on 802.11p (the MAC layer of DSRC) \cite{bazzi2014distributed,bazzi2016distributed}. Hagenauer et al.  investigated different options for leader election algorithms for VTL \cite{hagenauer2014advanced}. 

Other aspects of VTL scheme have also been studied. For example, Ferreira et al.  studied the impact of VTL on carbon emissions \cite{ferreira2012impact}. The simulation results show that VTL will lead to a 20\% reduction on $CO_2$ emissions under high-density traffic. Viriyasitavat et al.  explore the required public policy aspects to accelerate the adoption of VTL \cite{viriyasitavat2013accelerating}. In this work, the authors propose to have dedicated routes for VTL. Such a scheme was later on  studied in more detail by Tonguz et al. \cite{tonguz2014implementing}. The study is based on game-theoretic arguments and proposes to allocate certain routes at designated times to the exclusive use of DSRC-equipped vehicles. Meanwhile, Nakamurakare et al.  report a VTL prototype device based on smartphone and wifi-direct technology that was tested on the campus of Carnegie Mellon University\cite{nakamurakare2013prototype}.

In this paper, we report:
\begin{enumerate}
\item A VTL prototype system based on DSRC radios;
\item A system design for implementing VTL scheme;
\item Promising preliminary results based on field trials conducted in Pittsburgh, that show the viability of VTL system in reducing commute time. The results can also be used as a benchmark for other DSRC based applications at intersections.
\end{enumerate}

\vspace{-0.1in}
\section{System Design}

\vspace{-0.05in}
\subsection{Components}
\begin{figure}[ht]
\centering
\includegraphics[width=3in]{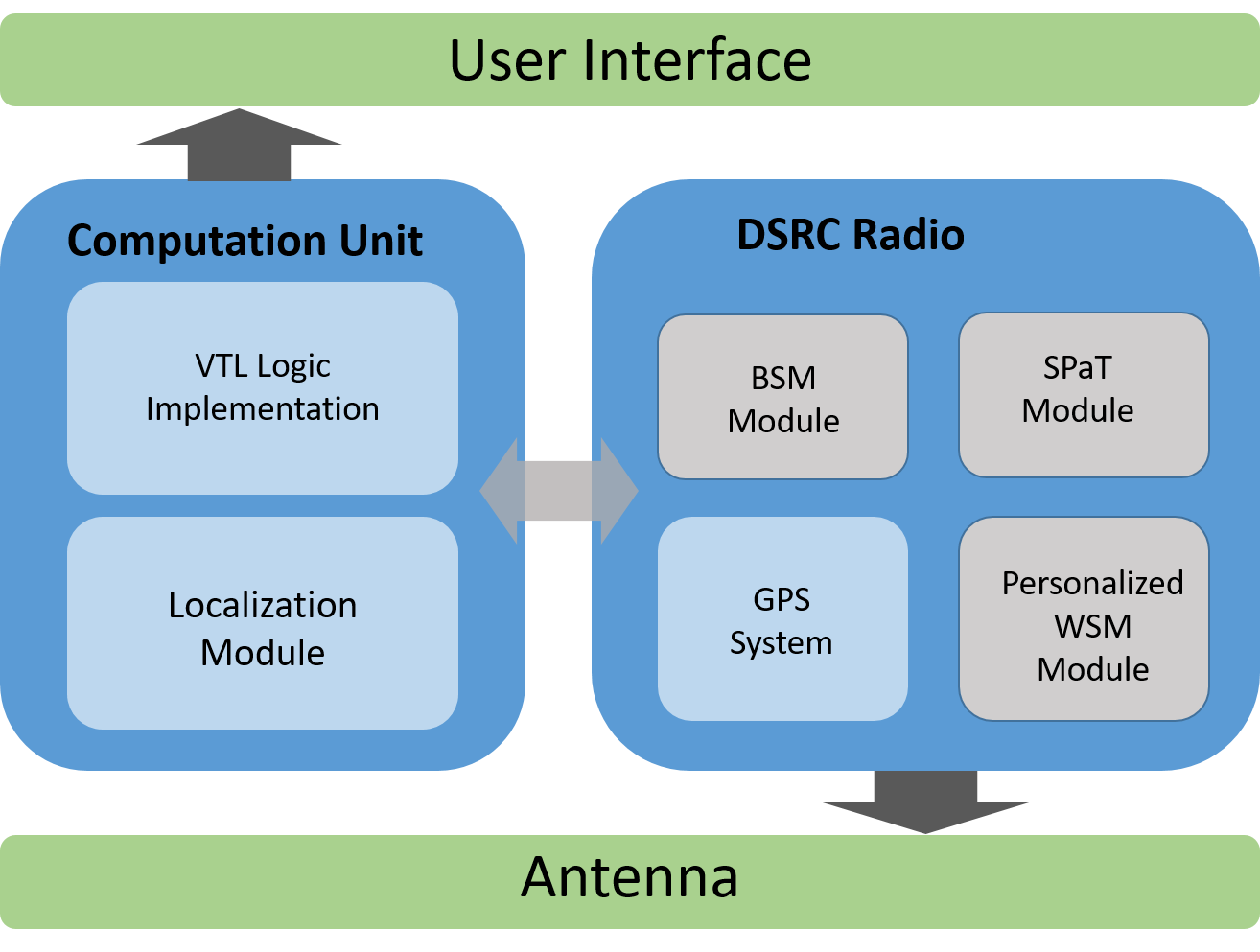}
\caption{Components of the prototype system}
\label{fig_components}
\vspace{-0.1in}
\end{figure}

Figure \ref{fig_components} shows the components of our prototype system. The system is composed of 2 major parts, the DSRC radio and Computational Unit. In general, the DSRC radio handles message transmission and reception from surrounding vehicles, and the Computational Unit makes a decision for the right-of-way in every vehicle approaching the intersection.

\subsubsection{DSRC Radio}
\label{SS: radio}

DSRC radio is a short to medium range radio working in the 5.9 GHz band (5.850-5.925 GHz), which is allocated by the Federal Communications Commission (FCC) to be used for vehicle-related safety and mobility systems \cite{what, kenney2011dedicated}. 

\begin{figure}[htb]
\centering
\includegraphics[width=3in]{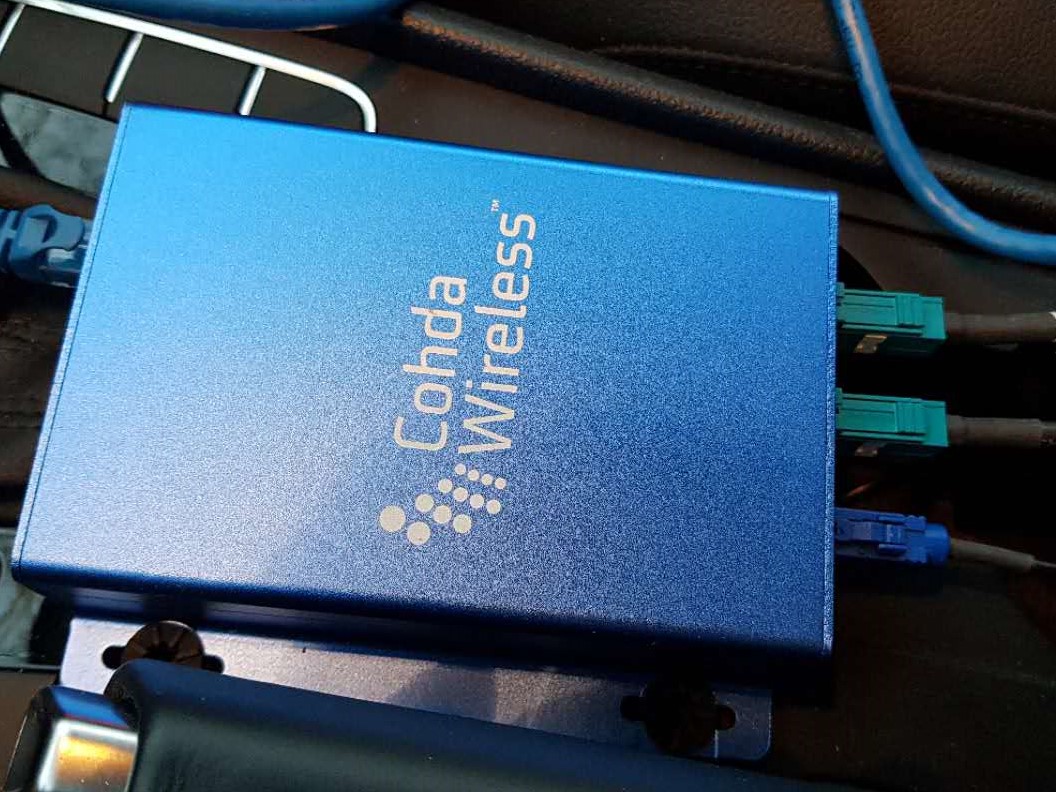}
\caption{DSRC radio unit used for our prototype system, manufactured by Cohda Wireless.}
\label{fig_radio}
\vspace{-0.1in}
\end{figure}

The DSRC On-Board Unit (OBU) is capable of broadcasting several types of messages specified by the SAE 2735 protocol \cite{sae20142735}. In the prototype design, we are mainly interested in Basic Safety Message (BSM) and Signal Phase and Timing  (SPaT) message.

BSM is a message type specified by SAE2735 protocol to be broadcast by each OBU and provides situational data to its surroundings. This message contains vehicle's current information including GPS coordinates, speed, and heading. By sensing and broadcasting BSM in real time, vehicles are aware of the status of their neighboring vehicles. Even though this message is designed mainly for safety applications, the message can be used for other applications as well. In our case, this message is used for discovering arrivals at the intersection on orthogonal approaches (streets).

SPaT message is yet another type of message specified by SAE 2735. The message provides traffic signal timing information to the surrounding vehicles. A SPaT message has a movement state field indicating the current traffic light phase. This message is typically broadcast by a Road Side Unit (RSU) to provide current traffic phase information to the traveler arriving at the intersection. In our application, instead of using RSU, we use an OBU to broadcast the current phase information. Details of this operation are provided in \ref{ss:vtl_logic}.

Other than the 2 standardized messages specified in SAE 2735 protocol, the prototype also generates some personalized WAVE Short Messages (WSM) for leader selection operation. These personalized messages from the application are only used during leader selection phases/handover phases (refer to section \ref{ss:vtl_logic}) and only comprise a very small proportion of all messages. Therefore, these messages will not affect the precious DSRC bandwidth in a major way.

Figure \ref{fig_radio} shows the DSRC OBU used in the prototype application, made by Cohda Wireless. The unit comes with embedded GPS system and a mag-mount DSRC+GPS antenna. 

In our prototype, we use the OBU's embedded GPS system for vehicle localization. In the actual deployment, for a high-end vehicle with a more accurate GPS system, the OBU has a choice of using the GPS signal that the vehicle provides.

\subsubsection{Computation Unit}
We use another computation unit that works with the DSRC radio to implement the VTL algorithm and decide the right-of-way for all vehicles approaching the intersection. For this specific prototype, we choose a Windows tablet with .net framework as our development platform. 

The application running on the computation unit is composed of two modules: a VTL logic module and a localization module. The localization module takes the GPS coordinates and provides geo-information, needed for the VTL logic module to make decisions. The VTL logic module then runs the VTL algorithm (refer to \ref{ss:vtl_logic}) based on this information.

In the prototype, we use a tablet as the computation unit. In the actual deployment, the computation unit can come in different forms, such as a smartphone application, or embedded in the vehicle's infotainment system.

\begin{figure}[htb]
\centering
\includegraphics[width=2.5in]{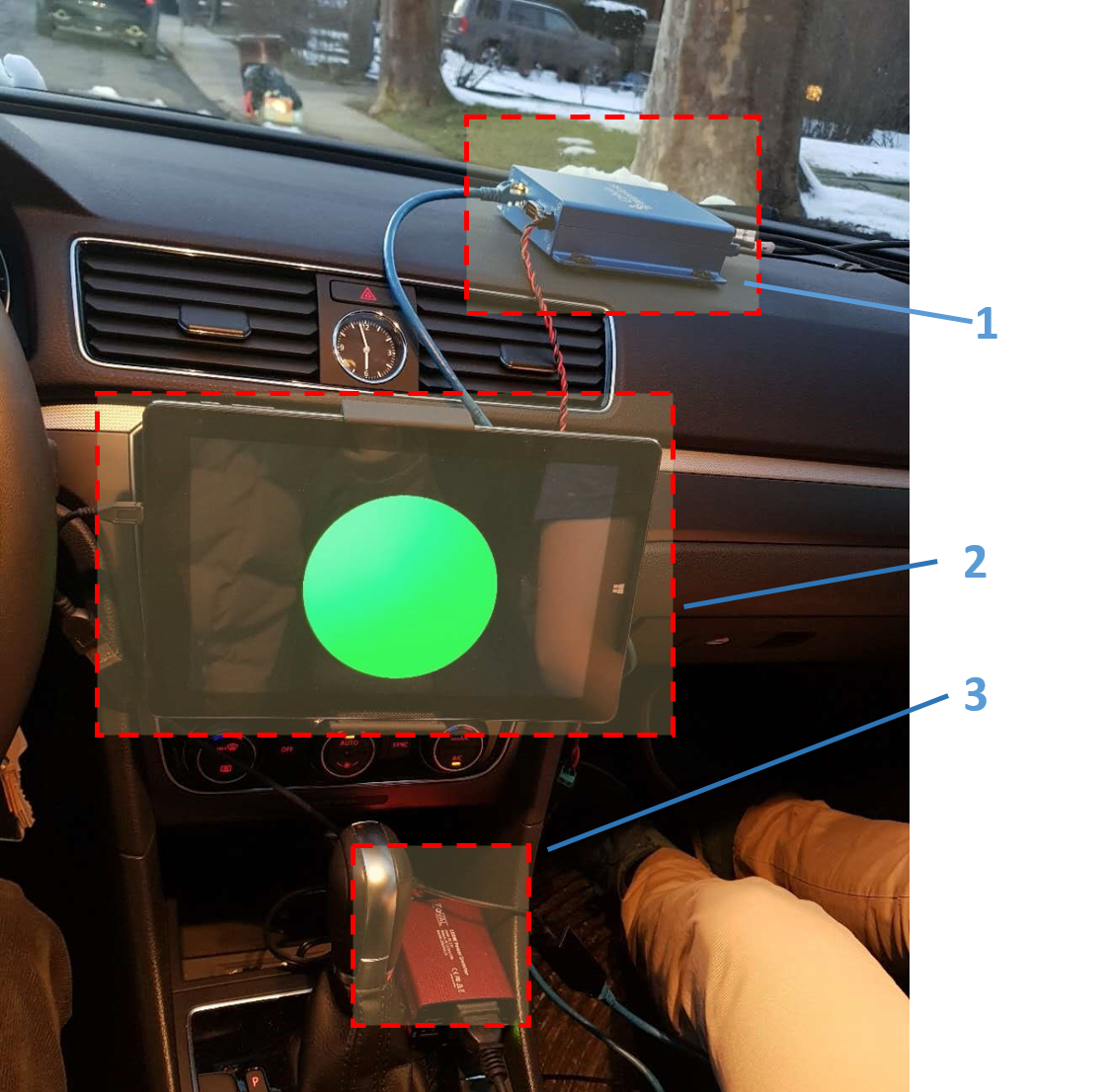}
\caption{A view of the VTL prototype inside a car}
\label{fig_carView}
\vspace{-0.1in}
\end{figure}

Figure \ref{fig_carView} shows the system inside the car. In this figure, we see the components mentioned above, including the DSRC radio (1), the Windows tablet which serves both as the computation unit and user interface (2), and a power adapter (3). The system in the figure is giving the vehicle a green light. 
\subsection{VTL Logic}
\label{ss:vtl_logic}

The VTL algorithm has been discussed in \cite{ferreira2010self} at a high level. To briefly recap, the principle of operation is as follows:
\begin{enumerate}
\item \label{VTLStep_sensing} \emph{Sensing:} While vehicles are approaching the intersection, they start to detect other vehicles approaching the same intersection.
\item \label{VTLStep_election} \emph{Leader Election:} If a conflict is detected, vehicles select a leader which temporarily serves as a traffic light.

\item \label{VTLStep_broadcast} \emph{Broadcast:} The leader broadcasts traffic light information. The leader will give red phase to its own lanes and green phase to the lanes in the orthogonal direction.

\item \label{VTLStep_handover} \emph{Handover:} After the leader is elected, it decides how long each direction should receive the right-of-way. The time can be either a dynamic value based on current traffic needs or a fixed value. When the time is up, the leader will handover its leadership role to one of the vehicles in one of the orthogonal roads.

\item \label{VTLStep_release} \emph{Release:} When the leader doesn't detect any conflicting vehicles, it will give green to its own lane and then release the leadership functionality at the intersection. Now the intersection has no leader, and whenever there is a new conflict, vehicles will re-elect a leader, starting from step 1.
\end{enumerate}

To implement this algorithm using the radio system discussed in subsection \ref{SS: radio}, some more specific design choices are made. The \emph{Sensing} step is achieved by BSM, \emph{Leader Election} step is achieved using WSM, \emph{Broadcast} is achieved using SPaT, \emph{Handover} is achieved using WSM, and \emph{Release} can be achieved by stopping the broadcast.

\begin{figure}[hbt]
\centering
\includegraphics[width=3in]{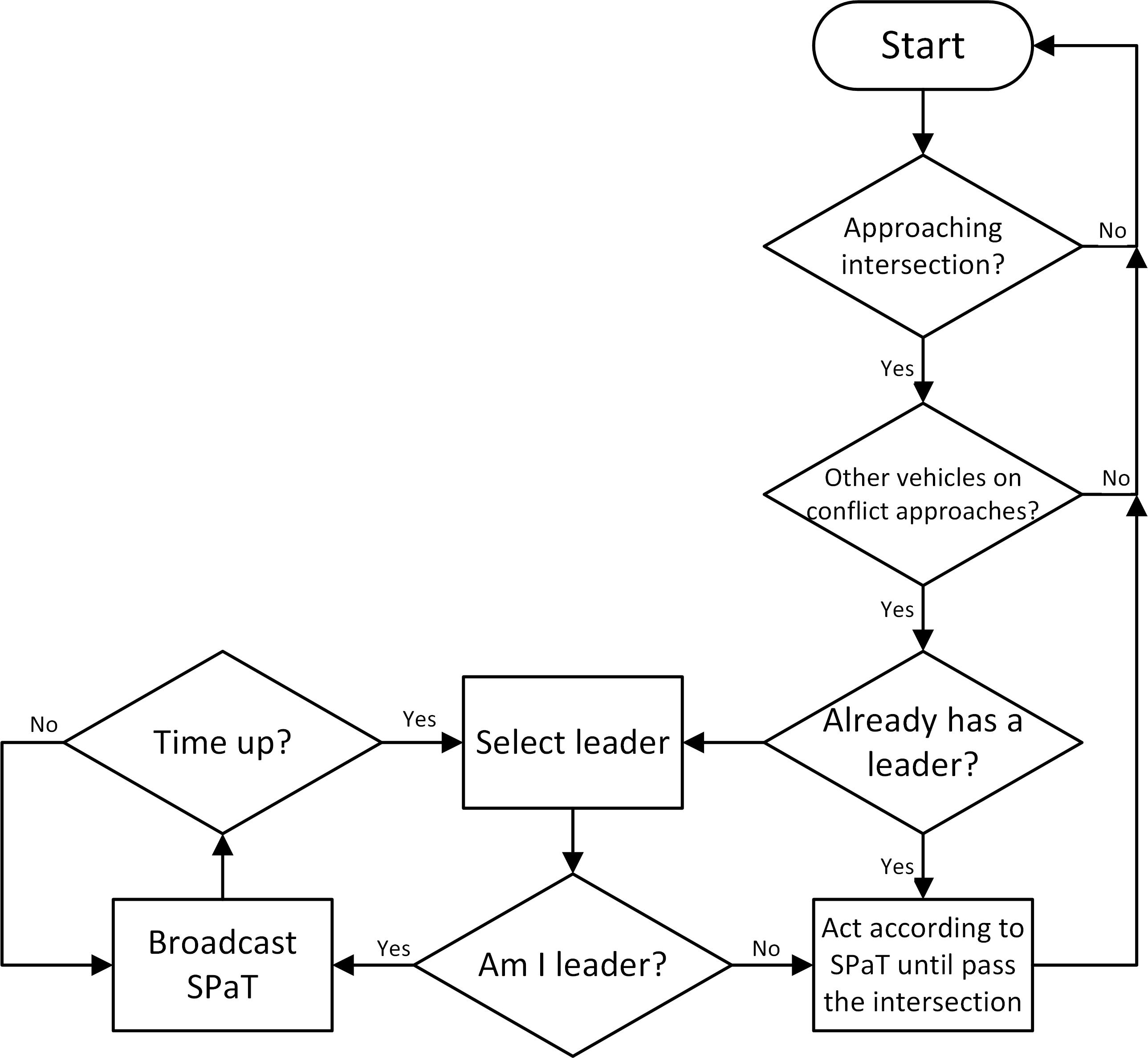}
\caption{Flowchart of the system implementation for VTL algorithm}
\label{fig_algo}
\vspace{-0.1in}
\end{figure}

Figure \ref{fig_algo} shows the flowchart of the system level implementation. Upon starting, the vehicle checks if it is approaching an intersection. If so, it detects other vehicles approaching the intersection from orthogonal approaches. If there is a competing arrival, the vehicle will check if there is a leader already at the intersection. If so, it will simply obey; otherwise, these vehicles will elect a leader to serve as a traffic light. After a time-up period, they will elect a different leader from the roads of next red light (\emph{Handover}). Observe that the \emph{Leader Election} and \emph{Handover} phase are both combined in the select leader block. When there is a leader already at the intersection, the select leader block will select another leader which is on one of the roads of the next red phase, thus achieving \emph{Handover} functionality.


\section{Performance}
In this section, we report some preliminary yet promising results using the field tests conducted in Pittsburgh, with the system introduced above. We present our results at two levels: the communication level results show the communications performance between vehicles at the intersection while the system level results show the end-user benefits.
\subsection{Communication Level}
\label{ss:comm_level}

\begin{figure}[ht]
\centering
\includegraphics[width=2.7in]{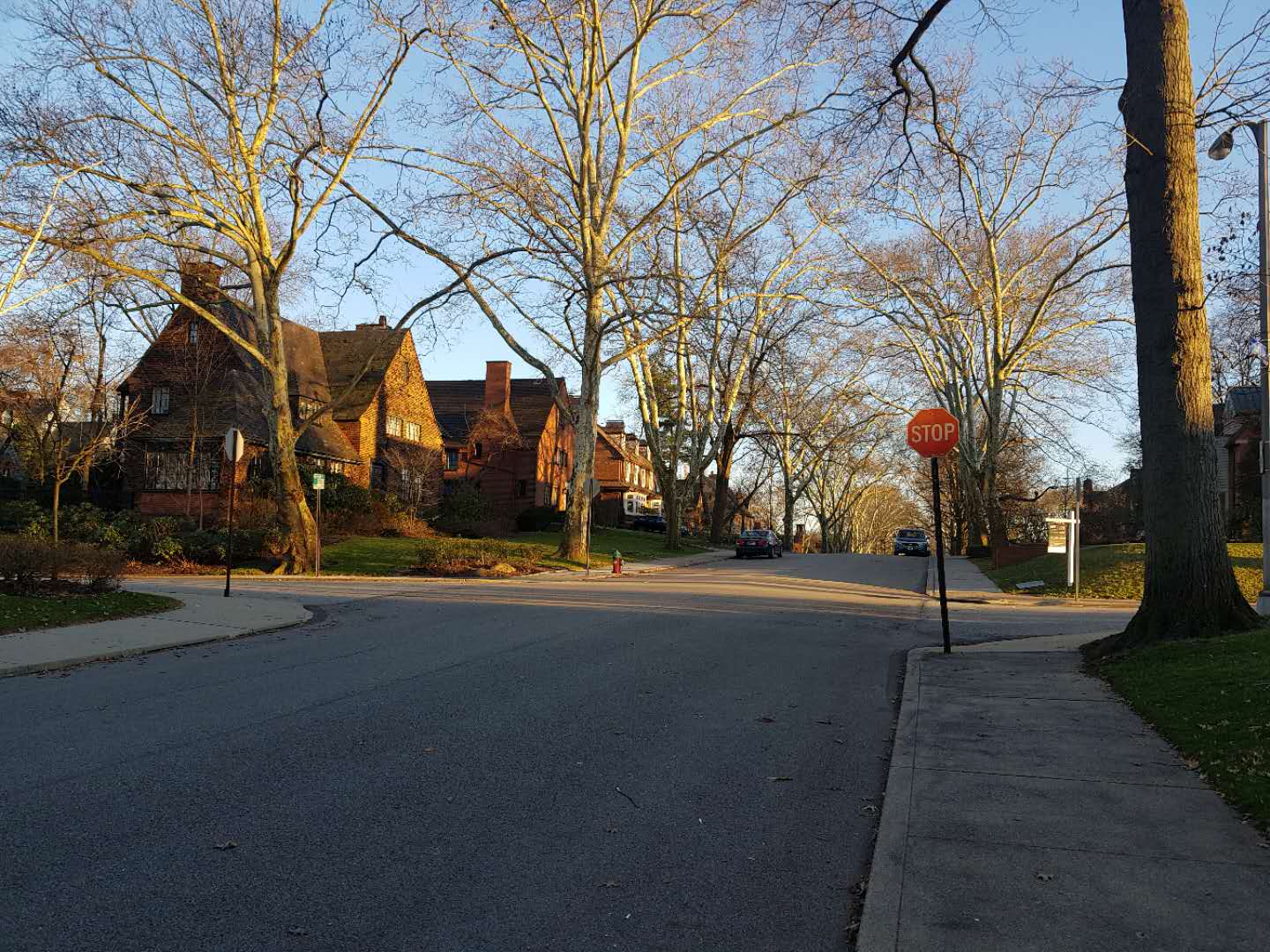}
\caption{The view of the intersection used for our field tests at the intersection of Bennington Avenue and Maynard Street in Squirrel Hill in Pittsburgh}
\label{fig_int}
\vspace{-0.1in}
\end{figure}
The communications between DSRC radios is the key factor that enables the whole functionality of the system. While there are multiple metrics to evaluate communications performance, in our application, the most important metric is Inter-Packet Gap (IPG), which is the time duration between two successfully received packets. In this context, it's the duration between two successfully received BSM. Since vehicles are using the received BSM to determine other vehicles' situation, IPG directly determines how often a vehicle updates its sensed data.

In our experiment, we park two vehicles on two perpendicular approaches at an intersection. One vehicle transmits BSM at a frequency of 10 Hz, the other vehicle receives the BSM and checks the IPG. We record the IPG when both vehicles are $50, 100, 150, 200, 250,$ and $300$ feet from the intersection, respectively. This shows the performance when both vehicles are approaching the intersection simultaneously. This is the scenario we are most interested in, since it's the situation most likely to create a conflict at the intersection. At each test point, more than 1000 BSMs are transmitted, and we report the average IPG from different distances to the intersection.

Figure \ref{fig_int} shows a picture of the intersection in Pittsburgh where we collect data. We place two vehicles on perpendicular approaches of this intersection. There are houses and trees between the transmitter and receiver, and this corresponds to a Non-Line-Of-Sight (NLOS) situation. Therefore, the results obtained here reflect real-world system performance.

\begin{figure}[hbt]
\centering
\includegraphics[width=3.5in]{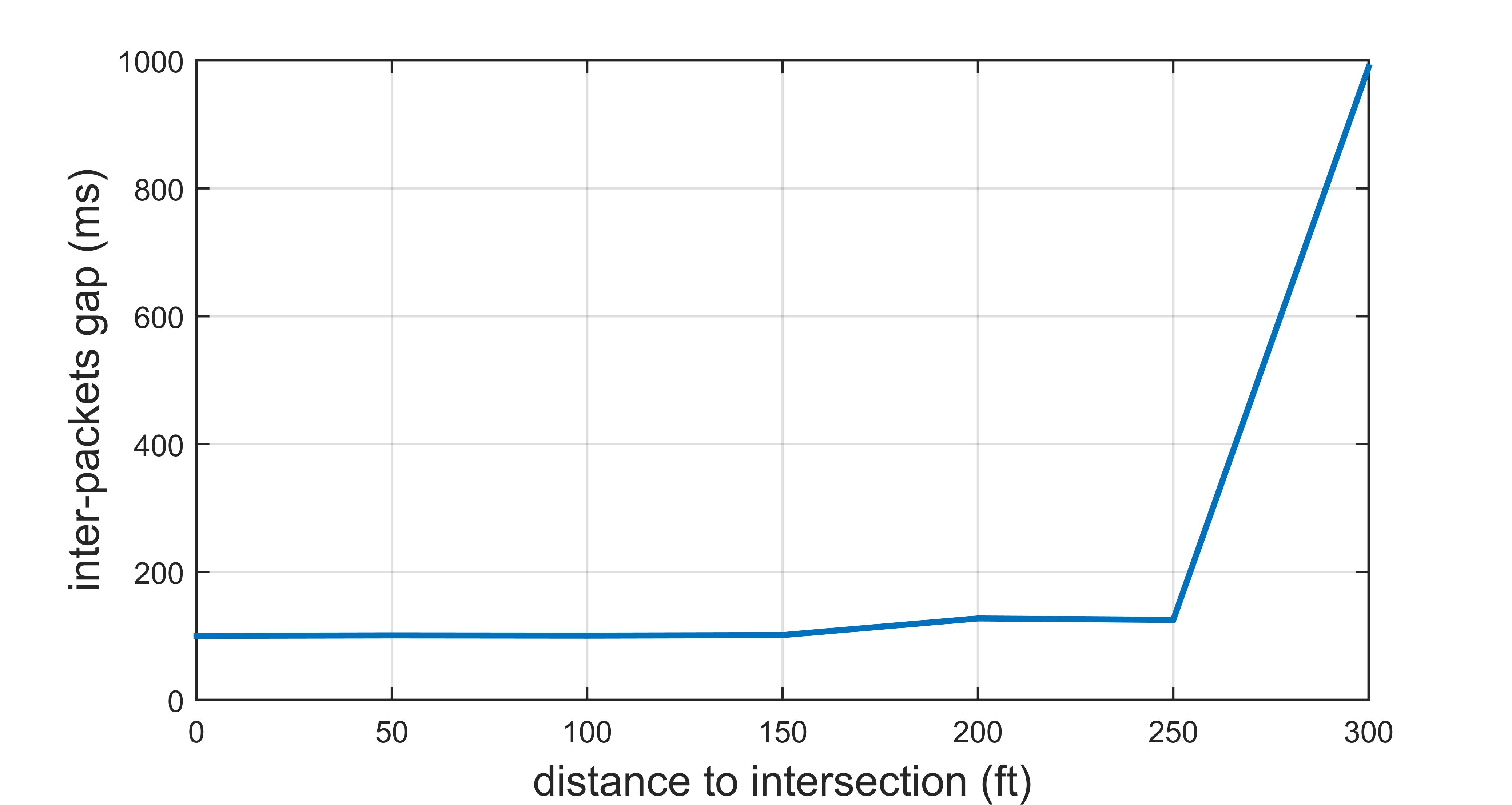}
\caption{Inter-Packets Gap at different distances to the intersection}
\label{fig_ipg}
\vspace{-0.1in}
\end{figure}

Figure \ref{fig_ipg} shows the results measured as a function of different distances to the intersection. The figure indicates that IPG is around 1 s when both cars are 300 feet from the intersection and as they get closer to the intersection the IPG starts to diminish. Observe that at 250 feet IPG has decreased to approximately 100 ms and it continues to operate at this performance level up until the vehicles are at the intersection.

From Figure \ref{fig_ipg} we can observe the performance of VTL system when both vehicles are approaching the intersection:
\begin{enumerate}
\item When both vehicles are 300 feet from the intersection, they start receiving messages and become aware of each other's existence.
\item The vehicles continue to approach the intersection and when they are within 250 feet (about 76 m) from the intersection, they start to receive messages from each other roughly every 100 ms. This means they can sense each other's status continuously, until they cross the intersection.
\end{enumerate}
The level of communication performance demonstrated in Figure \ref{fig_ipg} is sufficient for our application and provides a reliable basis for the designed VTL system to function properly.

\subsection{System Level Performance}
\label{ss:system-level}

The benefit of the VTL system can be evaluated for both the short and the long term. By short term, we refer to current or existing traffic trends whereby most people do NOT prefer routes with un-signalized intersections (as opposed to intersections with traffic lights). Consequently, the traffic density and the traffic flows at such intersections is low compared to routes with intersections equipped with traffic lights. Hence, commute time will decrease when one uses VTL since people do not need to stop if they are at the intersection first and also they do not need to stop at all if they are the only one at that intersection. In the short term, it is clear that VTL will eliminate the use of stop signs at un-signalized intersections. In the long term, since VTL can make un-signalized intersections behave like signalized intersections, the traffic will gradually re-distribute throughout the city. This will lead to an increase of  road capacity in the city. It is worth mentioning here that, in this paper, we only evaluate the short-term benefits of the VTL system. More specifically, we are interested in quantifying the benefits of replacing 4-way stop signs with VTL technology.

\begin{figure}[hbt]
\centering
\includegraphics[width=3in]{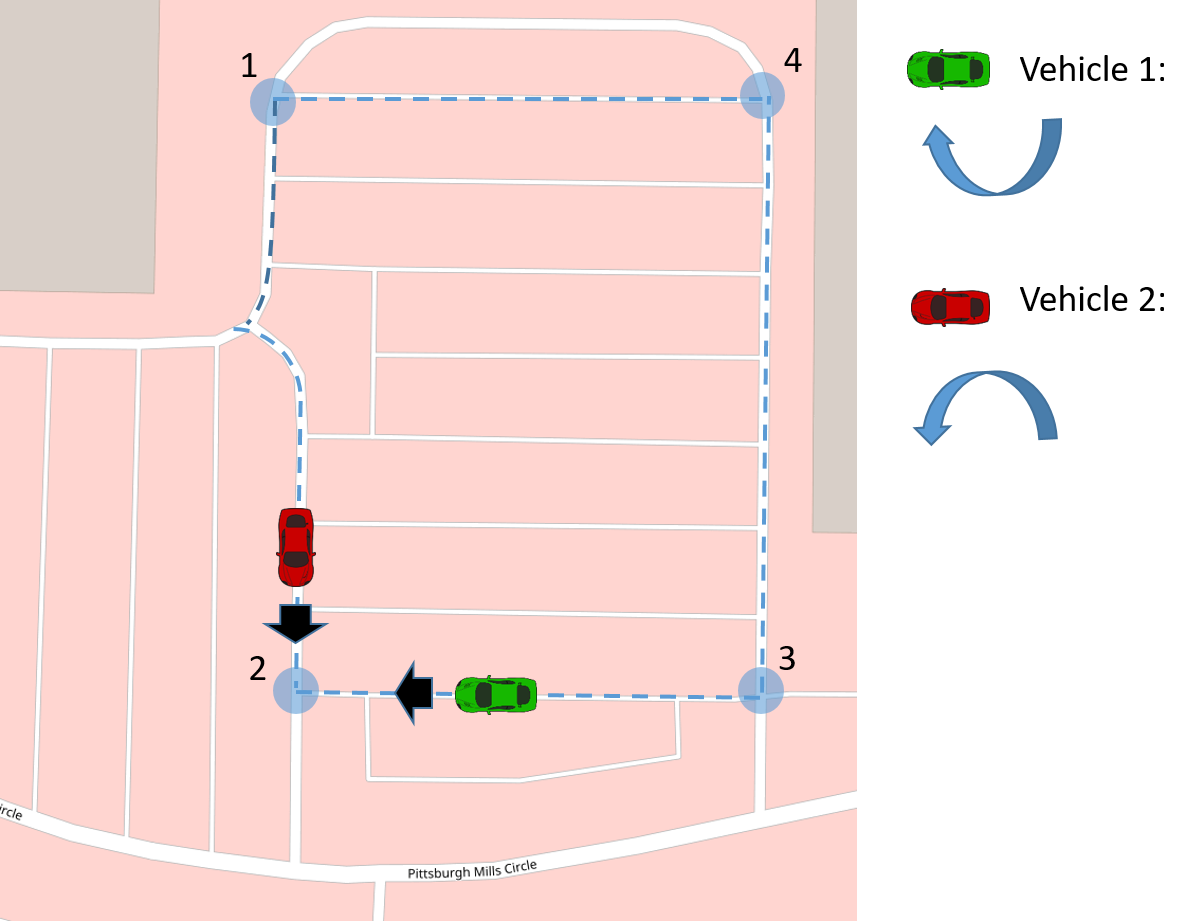}
\caption{Map and route used for the field test, map data from Open Street Map. The field tests were carried out in Pittsburgh Mills Gallery Parking Lot in Frazer Township, PA, in February 2018.
}
\label{fig_ft}
\vspace{-0.1in}
\end{figure}

The field tests are performed in an empty parking lot in Pittsburgh of size $300ft\times 160ft$. Figure \ref{fig_ft} shows the Open Street Map view and route used for the field test. We drive two vehicles in circles, as shown on the Open Street Map in Figure \ref{fig_ft}. The route of the two vehicles is plotted in blue dashed lines. As shown in the figure, vehicle 1 is driving clockwise, while vehicle 2 is driving counter-clockwise. Hence, they will possibly encounter each other at 4 corners of this rectangle, which we view as 4 intersections, as marked in the figure.

Since the rectangle has a length of 300 feet and a width of 160 feet, both of which are shorter than normal block size in the USA which is 410 feet, for the field tests, we reduce the car speed accordingly. During the filed tests we make sure that both vehicles are driving at 10 mph. We consider two different scenarios here:
\begin{enumerate}
\item Each intersection has a four-way stop sign.
\item VTL system is assumed at each intersection.
\end{enumerate}

For each intersection, both vehicles do 5 loops (which is equivalent to crossing 20 intersections) at the same time, one clockwise, the other counter-clockwise. We record the total time taken by both vehicles, under both scenarios.

\begin{table}[h]
\centering
\caption{Results of field test}
\label{tab-ft}
\begin{tabular}{|l|l|l|l|}
\hline
          & Stop sign & VTL  & Benefit \\ \hline
Vehicle 1 & 513 s      & 398 s & 22.4\%  \\ \hline
Vehicle 2 & 545 s      & 418 s & 23.3\%  \\ \hline
\end{tabular}
\end{table}

Table \ref{tab-ft} shows the results of this field test. We observe that for both vehicles, using VTL reduces the total travel time by 20\% compared to the use of stop signs at each intersection. Though the results are preliminary, it shows that the designed VTL system is promising and will be able to reduce the commute time by 20\% at those intersections with stop signs.

\section{Discussions}
The results reported in Table \ref{tab-ft} is the overall commute time of 5 loops, equivalent to 20 intersections. This is a typical number of un-signalized intersections a daily commuter might go through in a single trip during rush hours. The results show the performance difference, under sparse vehicle arrival rate, between a 4-way stop sign and VTL. 
Since an un-signalized intersection normally has a low vehicle arrival rate, the results correspond to the performance immediately after replacing 4-way stop signs with VTL (referred to 'short-term benefit' in subsection \ref{ss:system-level}). Namely, the average commute time for a commuter to cross these un-signalized intersections will be reduced by more than 20\% immediately after replacing 4-way stop signs.

After replacing 4-way stop signs with VTL, as time goes on, the traffic demand will shift asymptotically to Wardrop equilibrium \cite{Wardrop}. Since more and more vehicles will be using the roads with un-signalized intersections, the overall effective or used capacity of the roads in the area will increase (referred in subsection \ref{ss:system-level} as 'long-term benefit'). We envision more than 30\% reduction in commute time during this phase, according to the previous simulation results in \cite{ferreira2010self}.  

Further research is needed to address several important practical issues. One such issue is how to handle pedestrians  in the VTL scheme. One approach to solving this problem is to use a pedestrian-interactive RSU showing the current pedestrian phase, or a smartphone based app using Bluetooth Low Energy 5.0 (BLE 5.0). Meanwhile, the effect of NLOS situations should be carefully considered as well. For some intersections with severe NLOS conditions, the VTL scheme for NLOS described in \cite{neudecker2012feasibility} could be implemented.

\section{Conclusions}
In this paper, we report a new prototype system of Virtual Traffic Lights based on DSRC technology. We provide detailed system and algorithm design, together with preliminary field test results. 

Our results show that VTL technology can indeed sense the ensuing conflict between vehicles at perpendicular lanes 250 feet away from the intersection, on a regular basis, at 100 ms intervals. This provides experimental evidence on how VTL technology can manage traffic at an intersection via V2V communications using DSRC radios.

We have also performed field tests in a parking lot in Pittsburgh, to compare the VTL's performance with existing stop signs, for quantifying the short-term benefit that VTL provides. The results show that VTL reduces the commute time by more than 20\% on routes with un-signalized intersections.







\bibliographystyle{IEEEtran}
\vspace{-0.05in}
\bibliography{IEEEabrv,reference}
%



\end{document}